\documentclass[aps,prb,twocolumn,floats,epsfig,pdflatex]{revtex4}
\usepackage{amsmath}
\usepackage{amsfonts}
\usepackage{graphicx}
\usepackage{amssymb}
\usepackage{amsbsy}
\usepackage{color}
\usepackage{ulem}

\newcommand{\beq}{\begin{equation}}
\newcommand{\eeq}{\end{equation}}
\newcommand{\bea}{\begin{eqnarray}}
\newcommand{\eea}{\end{eqnarray}}

\begin{document}

\title{Detecting prethermal Floquet phases of Rydberg atom arrays}
\author{Somsubhra Ghosh$^1$, Diptiman Sen$^2$, and K. Sengupta$^1$}
\affiliation{$^1$School of Physical Sciences, Indian Association for the
Cultivation of Science, Kolkata 700032, India \\
$^2$Center for High Energy Physics, Indian Institute of Science, Bengaluru 560012, India}
\date{\today}

\begin{abstract}

We study the prethermal Floquet phases of a two-dimensional (2D)
Rydberg atom array on a rectangular lattice in the presence of a
periodic drive with large drive amplitude. We derive an analytic,
albeit perturbative, Floquet Hamiltonian using Floquet perturbation
theory (FPT) which charts out these phases and shows that the
transition between them can be accessed by tuning the drive
frequency. Using both numerical exact diagonalization on finite-size
arrays and analytical first-order Floquet Hamiltonian derived using
FPT, we show that these prethermal Floquet phases and the
transitions between them can be detected by studying the dynamics of
equal-time density-density correlation functions of the Rydberg
atoms. Our analysis thus provides a simple way of detecting these
phases and associated transitions in this system; such a detection
can be achieved in standard experiments which we discuss.

\end{abstract}


\maketitle

\section{Introduction}
\label{intro}

Quantum systems involving ultracold atoms in optical lattices have
been the subject of intense theoretical and experimental studies in
recent years \cite{uc1,uc2,uc3,uc4,uc5,uc6,uc7,uc8}. The reason for
such interest in these systems stems from their ability to act as
emulators of strongly correlated models. Moreover, they allow us to
explore parameter regimes and quantum dynamics of the emulated
models which is usually impossible to access in standard laboratory
setups. A typical example is the emulation of the Bose-Hubbard model
using $^{87}{\rm Rb}$ atoms; this has led detailed theoretical and
experimental studies on both the superfluid-insulators transition
and non-equilibrium quantum dynamics of this model
\cite{uc1,uc2,bh1,bh2, bh3,bh4,bh5,bh6}. Another such example is the
emulation of the tilted Bose-Hubbard model which has led to
realization of translation symmetry broken ground states in these
systems \cite{uc6,uc7,tib1,tib2,tib3}.

More recently experimental systems involving ultracold Rydberg atoms
have been experimentally realized \cite{ryd1,ryd2,ryd3,ryd4}. These
atoms experience strong interatomic van der Waals interaction in
their excited state leading to Rydberg blockade with tunable
blockade radius \cite{rydbl1,rydbl2}. An array of such atoms in an
one-dimensional (1D) optical lattice, namely, a Rydberg chain, is
known to host both Ising and non-Ising quantum critical points
\cite{st1,fss1}; the signature of the associated phase transition
has been experimentally verified \cite{ryd3}. The non-equilibrium
dynamics of such atoms has also been theoretically studied
\cite{subir1,ks1,scar1,scar2,scar3,scar4, ks2,ks3}. Interestingly,
the violation of the eigenstate thermalization hypothesis (ETH)
\cite{eth1,eth2} for dynamics starting from a class of initial
states has been experimentally observed in these systems
\cite{ryd4}; this phenomenon has been explained by invoking the
existence of an atypical set of athermal mid-spectrum quantum
states, namely, quantum scars \cite{scar1,scar2,scar3,scar4}. The
presence of such scars in the eigenspectrum of the Floquet
Hamiltonian of a periodically driven Rydberg chain has also been
predicted \cite{ks2,ks3}.

A natural extension of the above-mentioned studies on Rydberg chains
is to investigate higher-dimensional Rydberg atom arrays. Such
arrays are expected to host a rich variety of quantum ground states
that have no 1D analogs. For the 2D square arrays, such studies have
predicted the presence of several translational symmetry broken
ground states with definite density-wave orders; these ordered
states are separated from the featureless disordered ground state
via a second-order phase transition \cite{subir2,subir3}. In more
complicated non-bipartite lattices such as the Kagome lattice, where
the atoms are designed to occupy the links of the lattice (or
equivalently sites of a ruby lattice), such atom arrays are
predicted to host a spin-liquid quantum ground state over a wide
parameter regime \cite{ashvin1}. More recently, phases of Rydberg
atoms on a 3D pyrochlore lattice have also been studied
\cite{pyro1}. However, the non-equilibrium dynamics of such atom
arrays has not been studied so far.

In this work, we shall study the prethermal Floquet phases of a
periodically driven Rydberg atom array arranged as a rectangular
lattice. The effective Hamiltonian of such an array can be described
in terms of two states on a site with coordinate $\vec r=(j_x,j_y)$.
The first of these is the ground state of the atoms; we shall denote
this by $|g_{\vec r}\rangle$. The other state is the Rydberg excited
state and is denoted by $|e_{\vec r}\rangle$. Using these as the
basis states at each site, we can write the effective Hamiltonian
of these atoms as \cite{ryd1,ryd2,ryd3,ryd4}
\begin{eqnarray}
H &=& \sum_{\vec r} \left(\Omega \sigma_{\vec r}^x -\frac{\Delta}{2}
\sigma_{\vec r}^z \right) + \frac{1}{2} \sum_{\vec r, \vec r'}
V(|\vec r-\vec r'|) \hat n_{\vec r} \hat n_{\vec r'}, \label{ham1}
\end{eqnarray}
where $\vec \sigma_{\vec r}$ are Pauli matrices in the space of
states described above, $\sigma_{\vec r}^x= |g_{\vec r}\rangle
\langle e_{\vec r}| + |e_{\vec r}\rangle \langle g_{\vec r}|$, and
$\hat n_{\vec r}= (1+\sigma_{\vec r}^z)/2$ is the Rydberg excitation
density at the site $\vec r$ of the lattice. Here $\Omega>0$ denotes
the coupling strength between the ground state and the Rydberg excited
state, $\Delta$ denotes the detuning which we shall assume to be
uniform throughout the array, and $ V(|\vec r-\vec r'|)= V_0/ |\vec
r-\vec r'|^6$ denotes the van der Waals interaction between two
Rydberg excitations with strength $V_0$. In what follows, we shall
drive the detuning parameter, $\Delta$, of the model periodically
with drive amplitude $\Delta_0$ and frequency $\omega_D= 2\pi/T$,
where $T$ is the time period. For the square-pulse protocol, the
drive term is given by
\begin{eqnarray}
\Delta(t) &=& \delta - \Delta_0 ~~{\rm for}~~ 0 \le t < T/2, \nonumber\\
&=& \delta + \Delta_0 ~~{\rm for}~~ T/2 \le t < T, \label{sqp}
\end{eqnarray}
and $\Delta (t+T) = \Delta (t)$,
while for the cosine drive protocol, we have
\begin{eqnarray} \Delta(t) &=& \delta + \Delta_0 \cos (\omega_D t).
\label{cosp} \end{eqnarray}
In this work, we shall restrict ourselves to the regime where
the drive amplitude is large: $\Delta_0 \gg \delta, \Omega, V_0$.

The main results that we obtain in this work are as follows. First,
using Floquet perturbation theory (FPT) which uses the inverse drive
amplitude as the small parameter \cite{fptd,fptt,fptr}, we obtain an
analytic, albeit perturbative, Floquet Hamiltonian for the driven
system for both the square-pulse (Eq.\ \ref{sqp}) and the cosine
(Eq.\ \ref{cosp}) protocols. Our analysis reveals several ordered
Floquet phases with distinct density-wave orders which are separated
from the disordered state by second-order critical points. We also
show that tuning the drive frequency allows us to tune the system
between these phases and through the critical points. Second, we
complement our results obtained from the analytical Floquet
Hamiltonian with that from numerical exact diagonalization (ED)
starting from $H$ (Eq.\ \ref{ham1}) with $\Delta(t)$ given by Eq.\
\ref{sqp}.Our study reveals the existence of an exponentially long
prethermal timescale in the large and intermediate drive amplitude
regime; the properties of the driven system up to this timescale is
well described by the analytic Floquet Hamiltonian. Third, using the
exact evolution operator obtained from ED, we compute the
density-density correlation function of Rydberg excitations of the
driven atom array after $n$ cycles of the drive. We show that such a
correlator exhibits qualitatively distinct behavior in the
density-wave ordered and the disordered Floquet phases; thus it
serves as an experimentally relevant marker for the Floquet phases
and the transitions between them. We demonstrate this explicitly for
the disordered to the star and the checkerboard ordered Floquet
phases (see Fig.\ \ref{fig1}); we find that the above-mentioned
correlation function displays distinct long-time behaviors in the
ordered and disordered phases as well as at the transition point
between them. Thus it can be used to distinguish between these
Floquet phases and also locate the transition between them. Finally,
we discuss experiments which can test our theory and discuss
possible extensions of our study in these systems.

The plan of the rest of the paper is as follows. In Sec.\
\ref{analytic}, we derive the analytic Floquet Hamiltonian using
FPT. This is followed by Sec.\ \ref{fphases}, where we compare its
prediction to numerical results obtained using ED and obtain the
phase diagram of the driven system. Next, in Sec.\ \ref{detstab}, we
compute the correlation function $C$, study its behavior in
different Floquet phases, and also discuss the stability of these
phases. Finally, we chart out experiments which can verify our
theory, discuss possible extensions of it in these systems, and
conclude in Sec.\ \ref{diss}.

\section{Analytic Floquet Hamiltonian}
\label{analytic}

In this section, we shall derive the analytic Floquet Hamiltonian
for both the cosine and the square pulse protocols using FPT. The
details of the FPT method can be found in Refs.~\onlinecite{fptd,fptt,fptr}.

To obtain the Floquet Hamiltonian we first rewrite $H(t)= H_0(t) +
H_1$, where
\begin{eqnarray}
H_0(t) &=& - ~\frac{\Delta(t)-\delta}{2} \sum_{\vec r} \sigma_{\vec
r}^z, \label{zeroh} \\
H_1 &=& H_{1a} + H_{1b}, \quad H_{1a} ~=~ \sum_{\vec r} \Omega
\sigma_{\vec r}^x, \nonumber\\
H_{1b} &=& - ~\sum_{\vec r} \frac{\delta}{2} \sigma_{\vec r}^z ~+~
\frac{1}{2} \sum_{\vec r, \vec r'} V(|\vec r-\vec r'|) \hat n_{\vec
r} \hat n_{\vec r'}. \nonumber \end{eqnarray}
This decomposition of $H(t)$ is made such that the term with the
largest amplitude, $\Delta_0$, is included in $H_0$. We have also
separated the terms in $H_1$ into those which commute ($H_{1b}$)
with $H_0$ and those which do not ($H_{1a}$).

Next, we construct the evolution operator $U_0(t,0)= {\mathcal T}_t
\exp[-i \int_0^t H_0(t') dt'/\hbar]$ (where ${\mathcal T}_t$ is the
time-ordering operator) corresponding to $H_0(t)$. For the
square-pulse protocol, this yields
\begin{eqnarray}
U_s^{(0)}(t,0) &=& e^{-i \Delta_0 t \sum_{\vec r} \sigma_{\vec
r}^z/2\hbar} ~~{\rm for}~~ 0 \le t < T/2, \label{sqpu0} \\
&=& e^{-i \Delta_0 (T-t) \sum_{\vec r} \sigma_{\vec r}^z/2\hbar}
~~{\rm for}~~ T/2 \le t < T, \nonumber \end{eqnarray}
while for the cosine protocol we obtain
\begin{eqnarray}
U_c^{(0)}(t,0) &=& e^{i \Delta_0 \sin (\omega_D t) \sum_{\vec r}
\sigma_{\vec r}^z/(2\hbar \omega_D)}. \label{cpu0} \end{eqnarray}
Note that for both protocols $U_0(T,0)=I$ where $I$ denotes the
identity matrix; hence the zeroth-order Floquet Hamiltonian is
$H_F^{(0)}=0$.

To find the first-order terms, we use standard perturbation theory
which yields \cite{fptr}
\begin{eqnarray}
U_{c(s)}^{(1)}(T,0) &=& -~ \frac{i}{\hbar} \int_0^T dt U_{c(s)}^{(0)
\dagger}(t,0) H_1 U_{c(s)}^{(0)}(t,0). \label{fptfirst}
\end{eqnarray} To evaluate $U_{c(s)}^{(1)}$, we first note that the
terms in $H_{1b}$ (Eq.\ \ref{zeroh}) that commute with
$U^{(0)}_{c/s}$ can be evaluated simply. This yields
\begin{eqnarray}
U_{c(s)}^{(1b)}(T,0) &=& -\frac{i T}{\hbar} H_{1b}, \nonumber\\
H_{F c(s)}^{(1b)} &=& \frac{i \hbar}{T} U_{c(s)}^{(1b)}(T,0) =
H_{1b}. \label{ffl1} \end{eqnarray}

To evaluate the contribution of $H_{1a}$ to $U_{c(s)}^{(1)}$, we
first note that $U_{c(s)}^{(0)}$ is diagonal in the Fock basis and
can be written as
\begin{eqnarray}
U_s^{(0)}(t,0) &=& e^{-i \Delta_0 t E_m/(2\hbar)} |m\rangle\langle m|\quad
t \le T/2 \nonumber\\
&=& e^{-i \Delta_0 (T-t)E_m/(2\hbar)} |m\rangle\langle m| \quad t
> T/2, \label{decom} \\
U_c^{(0)}(t,0) &=& e^{i \Delta_0 E_m \sin\omega_D t/(2\hbar
\omega_D)} |m\rangle\langle m|, \nonumber
\end{eqnarray}
where $|m\rangle$ denotes a Fock state with $m$ Rydberg excitations
(or equivalently $m$ spin-up sites) and $L^2-m$ atoms in their
ground state, and $E_m$ is the eigenvalue of $\sum_{\vec r}
\sigma_{\vec r}^z$ in the state $|m\rangle$. We note that these
states are degenerate since their energies do not depend on the
positions of the Rydberg excitations. In this picture it is easy to
see that $H_{1a}$ changes the number of such Rydberg excitations in
any state by $\pm 1$; thus $H_{1a} |m\rangle \sim |m+1\rangle +
|m-1\rangle$. Furthermore the energy differences between the states
$|m\rangle$ and $|m\pm 1\rangle$ are given by $\Delta E_m^{\pm} =
E_m - E_{m\pm 1} = \mp 2$. Using this, and after some standard
algebra detailed in Refs.~\onlinecite{ks3}, we find
\begin{eqnarray}
H_{Fc}^{(1a)} &=& \Omega J_0\left(\frac{2\lambda}{\pi}\right)
\sum_{\vec r} \sigma_{\vec r}^x, \label{hf1c} \nonumber\\
H_{Fs}^{(1a)} &=& \Omega \frac{\sin \lambda}{\lambda} \sum_{\vec r}
\left( \cos \lambda \sigma_{\vec r}^x - \sin \lambda \sigma_{\vec
r}^y \right), \label{hf1s} \end{eqnarray}
where $\lambda = \Delta_0 T/(4 \hbar)$, and $J_0$ denotes the
zeroth-order Bessel function. The final first-order Floquet
Hamiltonian is given by
\begin{eqnarray} H_{F c(s)}^{(1)} = H_{F c(s)}^{(1a)}
+ H_{F c(s)}^{(1b)}. \label{hf1}
\end{eqnarray}

The expressions of $H_F^{(1)}$ for both the protocols suggest the
existence of special drive frequencies at which, for a given drive
amplitude $\Delta_0$, $H_{F}^{(1a)}$ vanishes. The frequencies
correspond to $\lambda= m \pi$, where $m$ is a non-zero integer, for
the square-pulse protocol and $\lambda= \pi \eta_m/2$ for the cosine
protocol, where $\eta_m$ denotes the value of the $m^{\rm th}$ zero
of $J_0$. They are given by
\begin{eqnarray}
\omega_m^{\ast} &=& \frac{\Delta_0}{2 m \hbar} ~~{\rm for~
square ~pulse~ protocol}, \nonumber\\
&=& \frac{\Delta_0}{\eta_m \hbar} ~~{\rm for ~cosine~ protocol}.
\label{magf}
\end{eqnarray}
At these frequencies $[H_F^{(1)}, \hat n_{\vec r}]=0$ leading to an
approximate emergent conservation of $\hat n_{\bf r}$. This
conservation is approximate since it is not respected by higher
order terms in the Floquet Hamiltonian.

To qualitatively understand the phases of $H_F^{(1)}$, we now
consider the regime where $V_0, \delta \ll \Delta_0$. In this regime
for $\omega_D \gg \omega_1^{\ast}$, the ground state of $H_F^{(1)}$
is expected to be similar to the disordered paramagnetic phase found
in Ref.\ \onlinecite{subir2}. In contrast at
$\omega_D=\omega_1^{\ast}$, the ground state of $H_F^{(1)}$
constitutes a density-wave ordered state whose precise nature
depends on the relative strength of $\delta/\Omega$ and
$V_0/\Omega$. \cite{subir2} Thus as we tune the drive frequency
towards $\omega_1^{\ast}$, we expect to find a second-order phase
transition between these phases. Also, in the regime of large
$\Delta_0$, the higher order corrections to $H_F^{(1)}$ are expected
to be small; thus such phases should persist as long-lived
prethermal phases of the driven system. We shall explore these
phases in detail in Sec.\ \ref{fphases} and their detection in Sec.\
\ref{corr}.

Before ending this section, we note that the effect of having
$V_0 \gg \delta, \Omega$ is to preclude Rydberg excitations on the
neighboring
sites of the lattice. In this regime, it is possible to obtain a
slightly modified form of the Floquet Hamiltonian which supports
similar phases. Such a prohibition can be implemented by using a
local projection operator
\begin{eqnarray}
P_{\vec r}= (1-\sigma_{\vec r}^z)/2 \label{proj1}
\end{eqnarray}
as shown in Ref.\ \onlinecite{ryd1}. In this regime, the projected
Hamiltonian is given by \cite{ryd1,st1,fss1}
\begin{eqnarray}
H_p(t) &=& \sum_{\vec r} \left(\Omega \tilde \sigma_{\vec r}^x
-\frac{\Delta(t)}{2} \sigma_{\vec r}^z \right) + \frac{1}{2}
\sum'_{\vec r, \vec r'} V(|\vec
r-\vec r'|) \hat n_{\vec r} \hat n_{\vec r'} \nonumber\\
\tilde \sigma_{\vec r}^x &=& P_{j_x-1,j_y} P_{j_x,j_y-1}
\sigma_{j_x, j_y}^x P_{j_x+1,j_y} P_{j_x,j_y+1}, \label{ham2}
\end{eqnarray}
where $\sum'$ denotes a sum over sites where $\vec r$ is not a nearest
neighbor of $\vec r'$. Note that $\tilde \sigma^x_{\vec r}$ can
create a Rydberg excitation at site $\vec r$ only if all its
neighbors are in their ground states.

We can carry out an exactly similar perturbative analysis, charted
out earlier in this section, starting from $H_p(t)$. The computation
involved is almost identical to that used to obtain $H_F^{(1)}$ and
we do not repeat it here. Such an analysis yields the Floquet
Hamiltonians for the continuous and the square pulse protocols for
the projected case.
\begin{eqnarray}
H_{F c}^{p(1)} &=& \Omega J_0\left(\frac{2\lambda}{\pi}\right)
\sum_{\vec r} \tilde \sigma_{\vec r}^x +\frac{1}{2} \sum'_{\vec r,
\vec r'} V(|\vec r-\vec r'|) \hat n_{\vec r} \hat n_{\vec r'} \nonumber\\
&& -\frac{\delta}{2}\sum_{\vec r}\sigma_{\vec r}^z\nonumber\\
H_{F s}^{p(1)} &=& \Omega \frac{\sin \lambda}{\lambda} \sum_{\vec
r} \left( \cos \lambda \tilde \sigma_{\bf r}^x - \sin \lambda
\tilde \sigma_{\vec r}^y \right) \nonumber\\
&& +\frac{1}{2} \sum'_{\vec r, \vec r'} V(|\vec r-\vec r'|) \hat
n_{\vec r} \hat n_{\vec r'} -\frac{\delta}{2}\sum_{\vec
r}\sigma_{\vec r}^z. \label{projfl}
\end{eqnarray}
We note that the phases of $H_F^{p(1)}$ are qualitatively similar to
those of $H_F^{(1)}$; in particular, we can still tune the drive
frequency towards $\omega_1^{\ast}$ to obtain density-wave phases.
The numerical advantage provided by $H_F^{p(1)}$ comes from the fact
that the dimension of its Hilbert space, ${\mathcal D}_p \sim
1.503^{L^2}$ \cite{baxter}, grows slowly with system size $L^2$
compared to its counterpart ${\mathcal D}$ for $H_F^{(1)}$,
${\mathcal D} \sim 2^{L^2}$. We shall use this fact while dealing
with numerical analysis of the Floquet phases in subsequent
sections.

\section{Prethermal Floquet phases}
\label{fphases}

\begin{figure}
\rotatebox{0}{\includegraphics*[width=\linewidth]{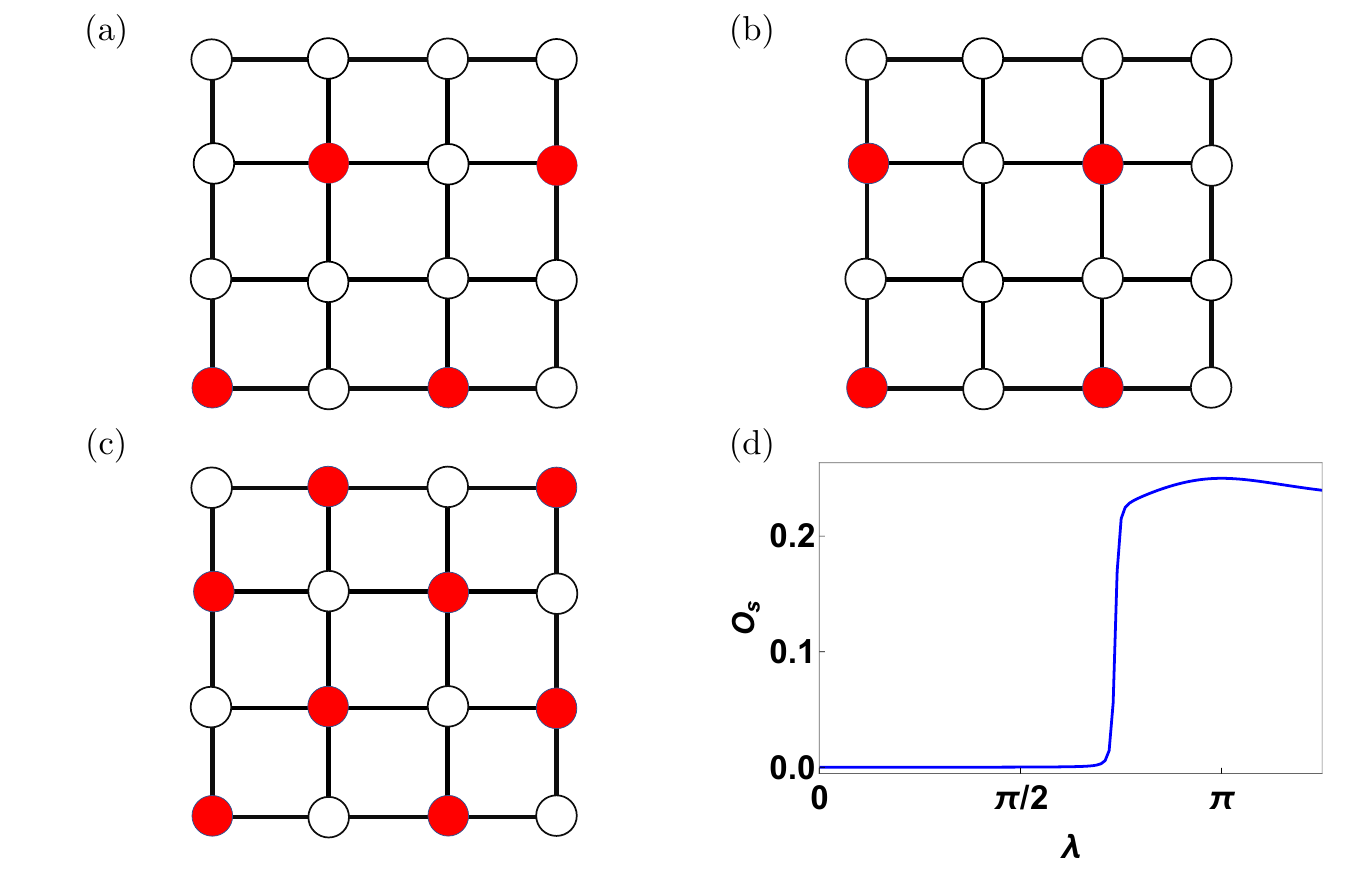}}
\caption{Schematic representations of the (a) star, (b) striated, and
(c) checkerboard phases. The red circles indicate sites with Rydberg
excitations while the white ones represent atoms in their ground
state. (d) Plot of $O_s$ obtained using exact eigenstates of
$U(T,0)$ as a function $\lambda=\Delta_0 T/(4\hbar)$, where $T$ is
the time period of a square pulse (Eq.\ \ref{sqp}) for
$V_0=25\Omega$. The other parameters are $\delta=2 \Omega$,
$\Delta_0=100 \Omega$, $L_x=6$ and $L_y=4$.} \label{fig1} \end{figure}

In this section, we study the Floquet phases using both the
analytically obtained Floquet Hamiltonian (Eq.\ \ref{projfl}) and
the Floquet eigenstates obtained from exact numerical
diagonalization of $U(T,0)$. The numerical results presented in this
section will be obtained for the square-pulse protocol with large
$V_0$. We shall also use the constraint that two neighboring sites
cannot be simultaneously occupied by Rydberg excitations: $\hat
n_{\bf r} \hat n_{\bf r'} =0$ if ${\bf r}$ and ${\bf r'}$ are
nearest neighbors. This approximation, which becomes accurate at
large $V_0$, allows us to access larger system size; the validity of
this approximation will be discussed in detail in Sec.\ \ref{diss}.

For the square-pulse protocol and within the constrained subspace
mentioned above, we can write the evolution operator as $U(T,0) =
U_+(T,T/2) U_-(T/2,0)$ where
\begin{eqnarray}
U_-(t,0) &=& \exp[-i H_p(\Delta_-) t/\hbar], \nonumber\\
U_+(t,T/2) &=& \exp[-i H_p[\Delta_+](t-T/2)/\hbar], \label{upm}
\end{eqnarray}
where $\Delta_{\pm}= \delta \pm \Delta_0$ (Eq.\ \ref{sqp}).

To obtain the exact Floquet eigenstates, we first numerically
diagonalize $H_{p \pm} \equiv H_p(\Delta_{\pm})$. This allows us to
obtain its eigenvalues and corresponding eigenvectors: $H_{p \pm}
|q_{\pm}\rangle = E_{q_{\pm}} |q_{\pm}\rangle$. Using these we can
write the evolution operator as
\begin{eqnarray}
U(T,0) &=& \sum_{q_+, q'_+} {\mathcal U}_{q'_+ q_+} |q'_+\rangle
\langle q_+|, \nonumber\\
{\mathcal U}_{q'_+ q_+} &=& \sum_{p_-} e^{i(E_{q'_+}+ E_{p_-})T/(2\hbar)}
c_{q'_+ p_-}^{\ast} c_{q_+ p_-}, \label{resol1} \end{eqnarray}
where $c_{\alpha \beta} = \langle \beta|\alpha\rangle$ are the
overlap coefficients between eigenstates of $H_+$ and $H_-$. Using
the matrix elements ${\mathcal U}_{q'_+ q_+}$ (Eq.\ \ref{resol1}),
we can numerically diagonalize $U(T,0)$ to obtain its eigenvalues
and eigenfunctions
\begin{eqnarray}
U(T,0) |m\rangle &=& \Lambda_m(T) |m\rangle, \quad \Lambda_m(T)=
e^{i \theta_m(T)}, \label{ueigen} \end{eqnarray}
where the form of the eigenvalues $\Lambda_m$ follows from the
unitary nature of the evolution operator. We note that the $|m\rangle$'s
are also eigenstates of the exact Floquet Hamiltonian $H_F$
within the constrained subspace; their eigenvalues which correspond to
the Floquet quasienergies are given by
\begin{eqnarray}
\epsilon_m &=& \arccos[{\rm Re}\Lambda_m(T)]\hbar/T. \label{eigen}
\end{eqnarray}
In the limit of high drive frequency where $T$ is sufficiently
small, all the eigenvalues fall within the first Floquet Brillouin
zone: $-\pi\hbar/T \le \epsilon_m \le \pi\hbar/T$. In this case, one
can meaningfully order the quasienergies $\epsilon_m$; the lowest
$\epsilon_m$ corresponds to the Floquet ground state and
characterizes the Floquet phase. For lower drive frequencies, the
quasienergies are no longer restricted within the first Floquet
Brillouin zone; in this regime, they can be folded back using the
standard reduced zone scheme \cite{fptr}. However, this makes it
impossible to order them by their magnitude. In this section, we
shall work in the high drive frequency regime where the eigenvalues
can be ordered.

To characterize the properties of these eigenvalues, we now define
the order parameters corresponding to various ordered phases of this
model \cite{subir2,subir3}. These ordered states, namely, the star,
the striated and the checkerboard states, are schematically sketched
in Figs.\ \ref{fig1} (a), (b), and (c) respectively. To characterize
such orders, we label the sites of a $L_x \times L_y$ lattice by an
integer
\begin{eqnarray}
j &=& (j_x-1)+(j_y-1)L_x  \label{marker}
\end{eqnarray}
where $1 \le j_y \le L_y$ is the row index of the array and $1 \le
j_{x} \le L_{x}$ is the $x$ coordinate of the site. We then define
an operator
\begin{eqnarray}
\hat O_c &=& \frac{1}{L} \sum_{j} (-1)^{j+[j/L_x]} (\hat n_j-1/2),
\label{ocexp}
\end{eqnarray}
where $[x]$ denotes the largest integer smaller than or equal to $x$
and $L= L_x L_y$. It is straightforward to see that $\langle \hat
O_c\rangle =1/2$ for the checkerboard phase, $1/4$ for the striated
phase and $0$ for the star phase. In contrast, the operator $\hat
O_s$ defined as
\begin{eqnarray}
\hat O_s &=& \frac{1}{L} \sum_{j} (-1)^{j+[j/(2L_x)]} (\hat n_j-1/2)
\label{osexp}
\end{eqnarray}
vanishes for the checkerboard and striated phase; for the star phase
$\langle \hat O_s \rangle = 1/4$. A representative plot showing
behavior of $O_s$ across the transition from the paramagnetic to the
star phase is shown in Fig.\ \ref{fig1} (d); we find that such a plot
indicates a transition around $\lambda \simeq 3\pi/4$ for $V_0=25
\Omega$, $\delta=2 \Omega$ and $\Delta_0=100\Omega$. The behavior of
$O_c$ for transition from disordered to checkerboard or disordered
to striated phases are similar.

\begin{figure}
\rotatebox{0}{\includegraphics*[width=\linewidth]{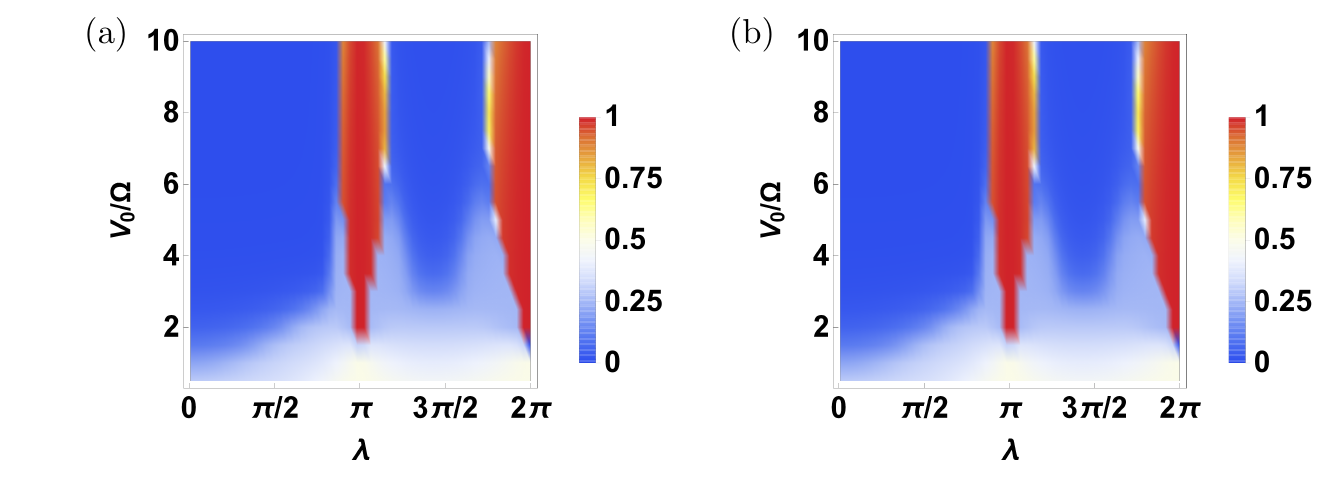}}
\caption{(a) Plot of $O_0$ obtained using eigenstates of $H_{F
s}^{p(1)}$ as a function of $V_0$ and $\lambda=\Delta_0 T/(4\hbar)$,
where $T$ is the time period of a square pulse (Eq.\ \ref{sqp}). The
dark red region corresponds to the star order for which $O_0 \simeq
1$, the white region corresponds to the checkerboard order for which
$O_0 \simeq 1/2$, and the light blue denotes the striated order for
which $O_0 \simeq 1/4$. The dark blue region represents a disordered
phase for which $O_0=0$. For these plots, $\delta=0.75 \Omega$,
$\Delta_0=100 \Omega$, $L_x=6$ and $L_y=4$. (b) Same as (a) but
obtained using exact numerical diagonalization of $U(T,0)$ within
the projected Hilbert space. } \label{fig2}
\end{figure}

The above considerations prompt us to define an operator given by
\begin{eqnarray} \hat O_1 &=& \hat O_c + 4 \hat O_s \label{opdef}
\end{eqnarray}
which allows us to distinguish between all three phases: $\langle
\hat O_1 \rangle=1/2 ~(1/4)$ for checkerboard (striated) phase and
$1$ for the star phase. We note that all of these order parameters
vanish in the paramagnetic phase leading to $\langle \hat
O_1\rangle=0$; this allows us to use $\hat O_1$ to identify the
Floquet phases of $H_F$.

Next, to study the Floquet phases, we plot the expectation value
\begin{eqnarray} O_0 ~=~ \langle m_0 |\hat O_1|m_0\rangle, \label{opexp}
\end{eqnarray}
where $|m_0\rangle$ is the eigenstate corresponding to the lowest
$\epsilon_m$. The left panel of Fig.\ \ref{fig2} shows the plot of
$O_0$ obtained using $H_{Fs}^{p(1)}$ (Eq.\ \ref{projfl}) as a
function of $V_0/\Omega$ and $\lambda=\Delta_0 T/(4\hbar)$ in the
large drive amplitude regime ($\Delta_0=100 \Omega$). The right
panel of Fig.\ \ref{fig2} shows a similar plot obtained using exact
diagonalization of $U(T,0)$ as discussed earlier in this section; we
find the two plots to be qualitatively similar for all $V_0/\delta$,
when $V_0, \delta \ll \Delta_0$. This demonstrates the validity of
the FPT in this regime.

In the high drive frequency regime where $\lambda \ll 1$, the plot
reflects the presence of the paramagnetic phase (dark blue region) for
which $O_0=0$ for $V_0 \gg \delta=0.75 \Omega$. In contrast, for
$V_0 \sim \delta$, we find a smooth interpolation between the
disordered and the checkerboard phase (light blue region). Such an
interpolation is an artifact of using the constrained Hilbert space
in our numerics; understandably, this approximation holds only for
$V_0 \gg \delta$. The presence of the disordered phase at $V_0 \gg
\delta$ and $\lambda \ll 1$ is consistent with the result of the
first-order Magnus expansion for which the Floquet Hamiltonian is just
the time averaged value of $H(t)$ and is given by Eq.\ \ref{ham1} with
$\Delta \to \delta$. This model is known to have a paramagnetic
phase for $V_0 \gg \delta$ in this regime \cite{subir2}.

For $V_0 \gg \delta$, we also find clear second-order transitions
from the paramagnetic to the star phase as the drive time period $T$
is varied. This transition occurs around $\lambda \simeq \pi, 2 \pi$
where $\Omega_{\rm eff} = \Omega \sin \lambda/\lambda \ll \delta,
V_0$ (Eq.\ \ref{magf}). For drive frequencies corresponding to
$\lambda \simeq \pi, 2 \pi$ and $V_0 \sim \delta$, we find the
checkerboard phase (white region). By increasing $V_0$ and keeping
$\lambda \simeq n \pi$ where $n$ is an integer, we find a transition
from the checkerboard to the star phase. This transition is expected
to be first order since it is a transition between two phases with
distinct classical orders. The disordered phase is absent since
$\Omega_{\rm eff} \simeq 0$ for these drive frequencies.

For lower drive frequencies, where the Floquet quasienergies are no
longer restricted within the lowest Floquet Brillouin zone, we can
not order the eigenvalues. We however note that such ordered states
still exist at the special drive frequencies given by $\lambda=
n\pi$ as eigenstates of the Floquet Hamiltonian up to a prethermal
timescale. Furthermore, their presence leaves a detectable signature
in the correlation function of the systems as we shall discuss in
the next section.

\begin{figure}
\rotatebox{0}{\includegraphics*[width= \linewidth]{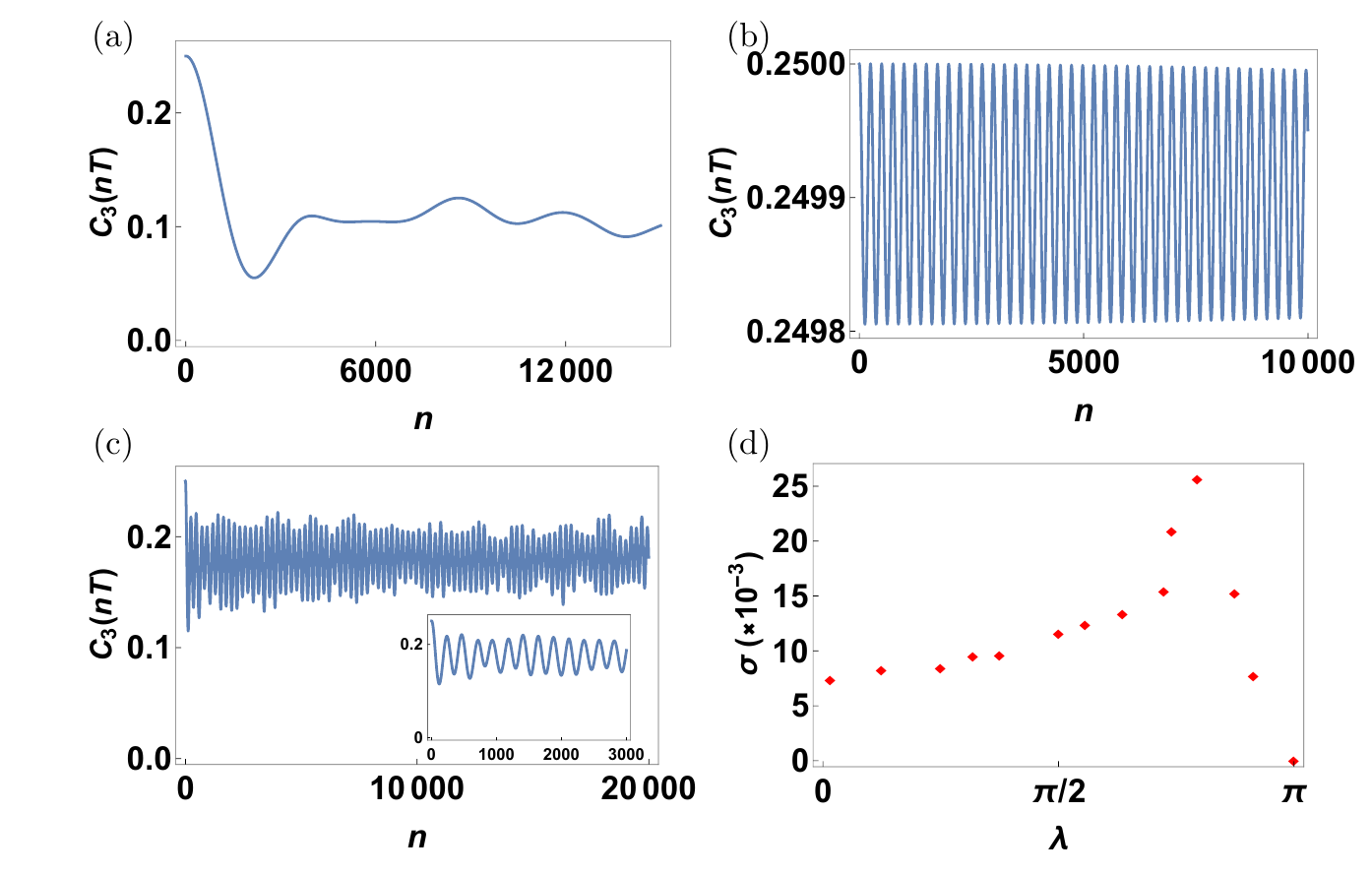}}
\caption{Plot of $C_3(nT)$ across the transition from the disordered
to the star phase. The plot shows $C_3(nT)$ as a function of the
number of drive cycles $n$ for (a) $\lambda=0.05$, (b) $\pi$, and
(c) $2.35$. The inset in panel (c) reflects the long-time
oscillations near the transition. The inset shows the details of
this oscillating behavior on a shorter time scale. Panel (d) shows
the fluctuation of the value of $C_3$ about its mean value as a
function of $\lambda$. For all plots $V_0=25 \Omega$, $\delta=2
\Omega$, and $\Delta_0=100 \Omega$. See text for details.}
\label{fig3} \end{figure}

\section{Detection and stability of the Floquet phases}
\label{detstab}

In this section, we shall first discuss the properties of the
correlation functions using which we can detect the Floquet phases.
This will be followed by a study of stability of these Floquet
phases and the extent of the prethermal regime as a function of the
drive amplitude. Throughout this section we shall work within the
projected Hilbert space as discussed earlier.

\subsection{Correlation functions}
\label{corr}

In this subsection, we show that the Floquet phases and the
transitions from the disordered paramagnetic to the star Floquet
phases can be detected via a study of correlation functions. A similar
detection scheme, as we shall discuss, is expected to hold for
transitions from the disordered to other ordered states. This is of
primary importance since, unlike equilibrium ground states, these
phases do not correspond to standard energy eigenstates states of a
many-body system and cannot be accessed via standard thermodynamic
measurements in experiments.

To this end, we compute the equal-time density-density correlation
function of the driven system given by
\begin{eqnarray}
C_3(nT) &=& \sum_{\vec r \vec a} \langle \psi(nT)| \hat n_{\vec r}
\hat n_{\vec r+\vec a} |\psi(nT)\rangle, \end{eqnarray} where $\vec
a$ is chosen so that $\vec r$ and $\vec r +\vec a$ form
third-nearest neighboring sites of a 2D rectangular lattice. Note
that the nearest-neighbor density-density correlation is identically
zero within the projected Hilbert space and the next-nearest
neighbor correlation is zero in the star phase and close to zero in
the disordered phase; thus $C_3$ represents the most local
correlation function with appreciable dynamical fluctuation.

The plot of $C_3$ is shown for three representative values of
$\lambda$ in Figs.\ \ref{fig3}(a), (b) and (c). In Fig.\
\ref{fig3}(a), $C_3$ is plotted as a function of $n$, the number of
drive cycles, for $\lambda=0.05$ and $V_0=25 \Omega$ and starting
from an initial Fock state $|\psi_0\rangle$ with star order
(sketched in Fig.\ \ref{fig1}) so that $\langle \psi_0|
C_3|\psi_0\rangle = 1/4$. The plot shows a rapid decay of the
correlator towards its diagonal ensemble value $\sim 0.1$ which is
in accordance with the prediction of ETH. The oscillations around
this value is a consequence of the finite system size. In contrast,
for $\lambda= \pi$, as shown in Fig.\ \ref{fig3}(b), $C_3$ remains
almost a constant showing very small oscillations around the initial
value. This is a consequence of the fact that $|\psi_0\rangle$ is
almost exactly an eigenstate of the exact Floquet Hamiltonian. We
note that this will happen as long as $|\psi_0\rangle$ is a
near-exact eigenstate of $U$ or, equivalently, of $H_F$; it need not
necessarily be its lowest-lying eigenstate. In between, near the
transition at $\lambda = 2.35$, we find that $C_3$ shows long-time
oscillatory behavior which is distinct from its counterparts shown
in Figs.\ \ref{fig3}(a) and (b). In particular, the oscillation
amplitudes are larger than their counterparts in the ordered phase;
also they are much longer-lived than what is found in the disordered
phase. This shows that $C_3$ can distinguish between the Floquet
phases and provides a straightforward tool for their detection. The
fluctuation of $C_3$ around its mean value at long-time is computed
as
\begin{eqnarray}
\sigma &=& \sqrt{\frac{1}{n_f-n_i}\sum_{n=n_i}^{n_f} C^2_3(nT) -\mu^2} \nonumber\\
\mu &=& \frac{1}{n_f-n_i}\sum_{n=n_i}^{n_f} C_3(nT).
\label{meanfluc}
\end{eqnarray}
with $n_i=3001$ and $n_f=10000$.

A plot of $\sigma$ as a function of $\lambda$ is shown in Fig.\
\ref{fig3}(d). We find that $\sigma$ indicates a clear peak at the
transition point $\lambda=\lambda_c \simeq 2.3$ indicating a sharp
increase in fluctuation of $C_3$ at the transition. It therefore
serves as a distinguishing feature of the transition from a
disordered to the star ordered phase.

\begin{figure}
\rotatebox{0}{\includegraphics*[width= \linewidth]{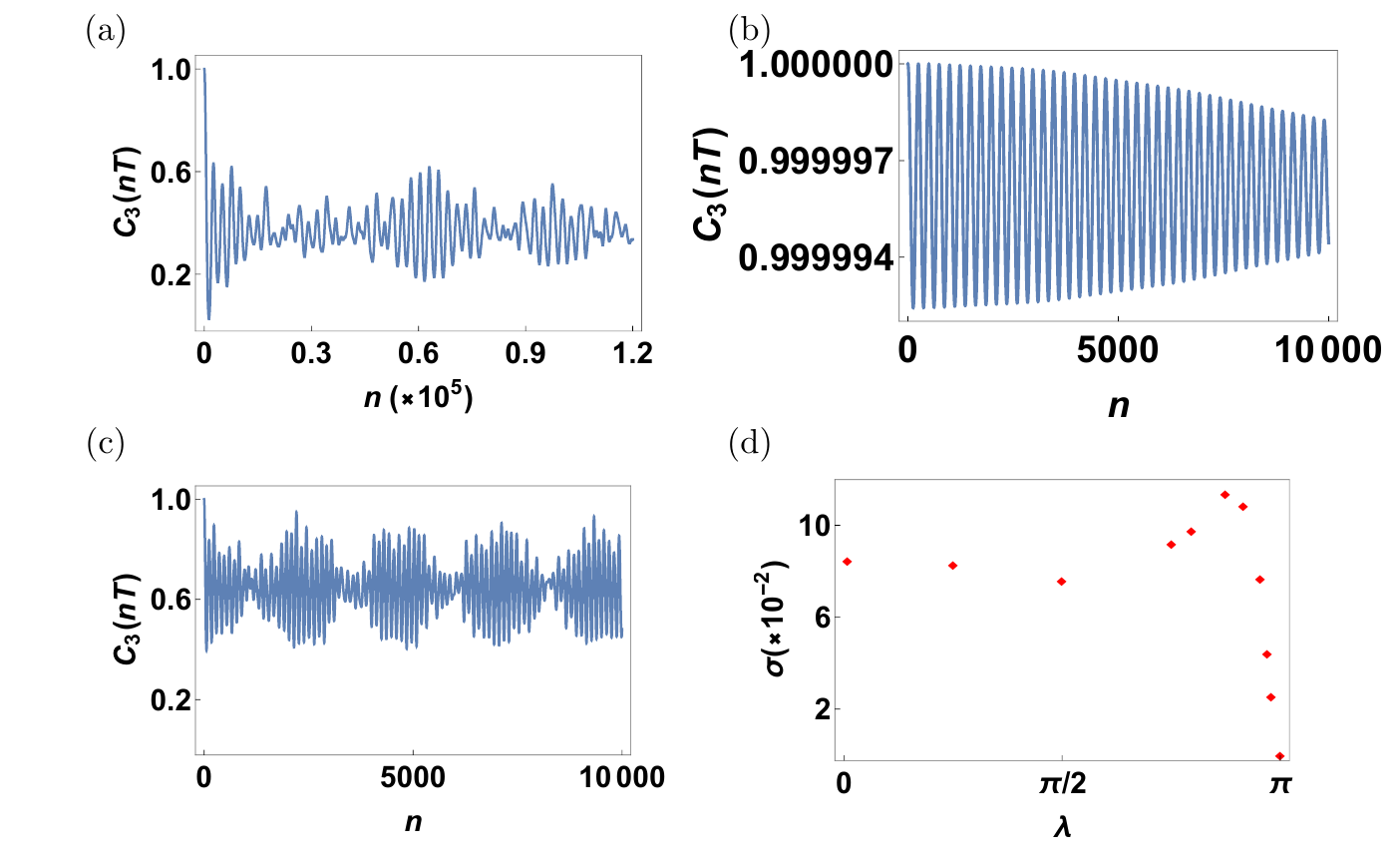}}
\caption{Plot of $C_3(nT)$ across the transition from the disordered
to the checkerboard phase. The plot shows $C_3(nT)$ as a function of
the number of drive cycles $n$ for (a) $\lambda=0.025$, (b) $\pi$,
and (c) $2.75$. Panel (d) shows the fluctuation of $C_3$ about its
mean value as a function of $\lambda$. For all plots $V_0=1.5
\Omega$, $\delta= \Omega$, and $\Delta_0=100 \Omega$. See text for
details.} \label{fig4}
\end{figure}

A similar signature in the behavior of $C_3(nT)$ is also noticed
when there is a transition between the checkerboard phase and the
disordered phase at a lower value of the interaction potential
$V_0$. This is shown in Fig. \ref{fig4} for $\delta=\Omega \text{
and } V_0=1.5\Omega$, where we choose a checkerboard ordered state
as our initial state $|\psi_0\rangle$, so that
$\langle\psi_0|C_3|\psi_0\rangle=1$. Away from the transition (Figs
\ref{fig4}(a) and (b)), the fluctuation of $C_3$ is comparatively
small, whereas close to the point of transition (Fig \ref{fig4}(c))
it peaks considerably. In particular, when $\lambda=\pi$ in Fig.
\ref{fig4}(c), the checkerboard state is almost an eigenstate of the
exact evolution operator, which is why the quantum fluctuations dip
to zero. In Fig. \ref{fig4}(d), we plot the average fluctuation of
$C_3(nT)$, $\sigma$ as a function of $\lambda$. The average is
computed after the initial transient dynamics have settled down. It
shows a peak around $\lambda=2.75$. We later show in Fig \ref{fig6}
that there is a phase transition from the disordered phase to the
checkerboard phase precisely at this point even when the full
Hilbert space is used instead of the projected subspace.

Before ending this section, we note that such correlators are
expected to show qualitatively similar behaviors across transitions
from a disordered to any other ordered Floquet phase provided that we
start from an initial Fock state which characterizes the order.
However, it is not expected to provide a signature of a transition
between two ordered phases; in this case, typically both the ordered
phases exist as eigenstates of the Floquet Hamiltonian across the
transition and the correlators do not evolve dynamically in either
of the phases provided that the initial state is one of the ordered
states.

\subsection{Stability of the Floquet phases}

In this section, we discuss the stability of such Floquet phases and
provide an estimate of the prethermal timescale over which such
phases are expected to exist. To this end, we first note that the
behavior of a driven ergodic system is expected to be described by a
local Floquet Hamiltonian only up to a finite, prethermal, timescale
$t_p$, where $\tau= t_p/T$. For $n > \tau$, the system is expected
to heat up to infinite temperature and can no longer be described by
a local Floquet Hamiltonian \cite{rigol1}. However, it is known, in
the context of Magnus expansion that $t_p \sim \exp[c \omega_D]$
(where $c$ is a constant of order 1 which depends on the system
details) in the high drive frequency limit \cite{saito1}. It can
thus be large leading to a long prethermal time over which the
Floquet phases are expected to be stable.

To estimate the prethermal timescale $\tau$ for the driven Rydberg
system, we plot $C_3(nT)$ as a function of number of drive cycles
$n$ at $\lambda=\pi$ for several representative values of
$\Delta_0/\Omega$. The value of $C_3(nT)$ obtained from $H_{F s}^{p
(1)}$ is a constant and equals $0.25$ for $\lambda= \pi$; at large
$\Delta_0$, such a constant value is also found for $C_3$ obtained
using ED as can be seen from Fig.\ \ref{fig3} (b). To characterize
the difference between the results obtained using ED and that from
$H_{F s}^{p (1)}$, we therefore study the deviation of $C_3(nT)$
from its constant value.

\begin{figure}
\rotatebox{0}{\includegraphics*[width= \linewidth]{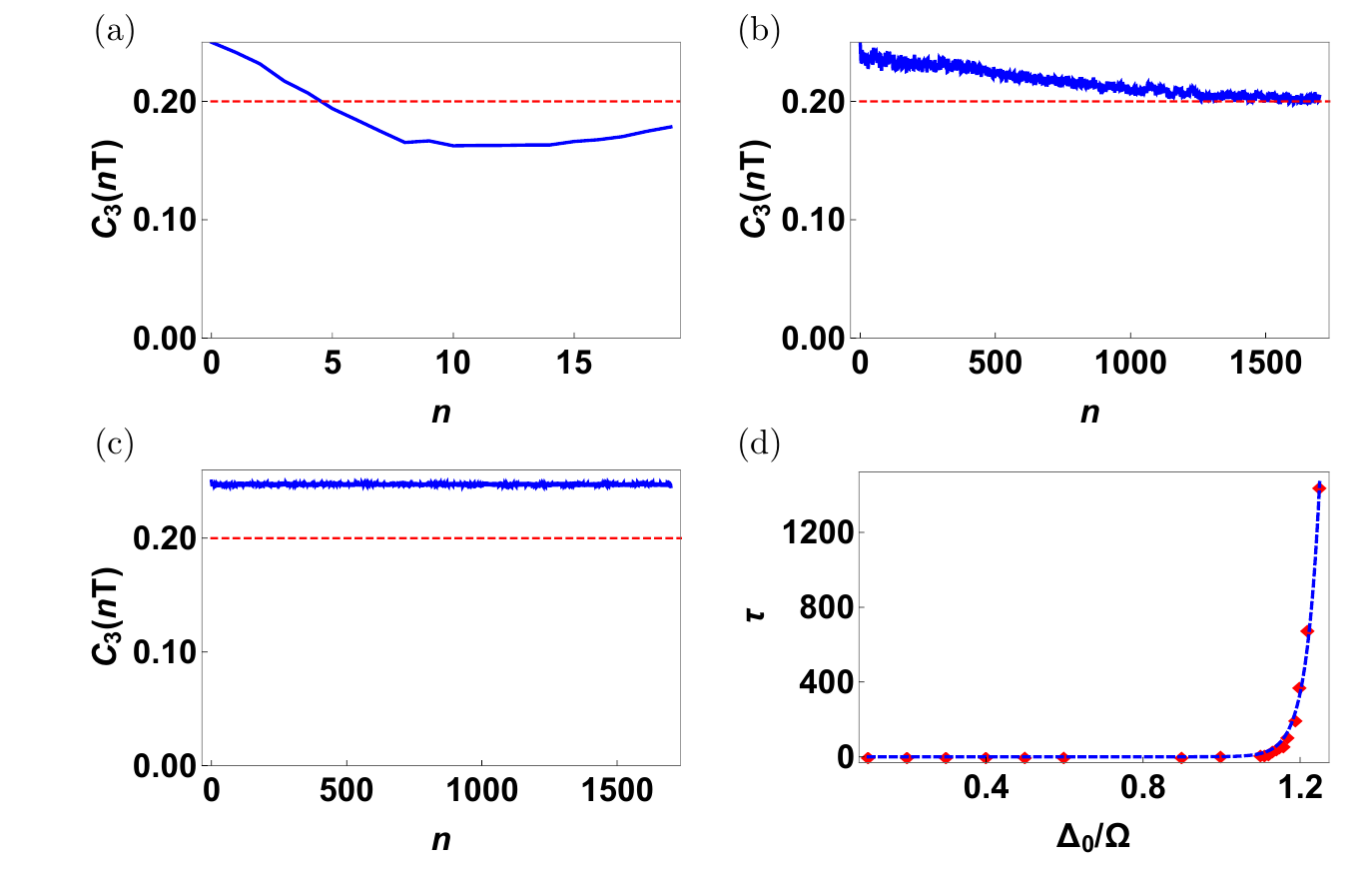}}
\caption{Plot of $C_3(nT)$ as a function of $n$ at $\lambda=\pi$ for
(a) $\Delta_0/\Omega=0.9$, (b) $1.25$, and (c) $1.45$. (d) Plot of
$\tau$, measured as the minimal number of cycles after which $C_3
\simeq 0.2$, as a function of $\Delta_0$ showing an exponential
growth of $\tau$ at large $\Delta_0$. For all plots $V_0=25 \Omega$
and $\delta=2 \Omega$. The red dotted lines in panels (a), (b) and
(c) indicate the line $C_3=0.2$. See text for details.} \label{fig5}
\end{figure}

The result of such a study is shown in Fig.\ \ref{fig5}. In Figs.\
\ref{fig5}(a), (b) and (c), we plot $C_3(nT)$, obtained using ED, as
a function of $n$ for $\lambda=\pi$. Fig.\ \ref{fig5}(a) shows such
a plot for a low drive amplitude $\Delta_0=0.9 \Omega$; we find that
$C_3(nT)$ deviates from its initial value within the first few drive
cycles. The time taken to achieve this deviation increases with
increasing $\Delta_0$ (Fig.\ \ref{fig5}(b) where $\Delta_0=1.25
\Omega$) and around $\Delta_0=1.45 \Omega$, $C_3(nT)$ remains fixed
at its constant value predicted by $H_{Fs}^{p(1)}$ for $n\gg 1500$
drive cycles.

From these plots, we can obtain a qualitative estimate of $\tau$.
Here we choose $\tau$ to be smallest number of drive cycles at which
$C_3(nT) \simeq 0.2$. The choice of $C_3(nT) \simeq 0.2$ as the
cut-off is motivated by the fact that the infinite-temperature
ensemble average of this correlator is close to $0.17$. A plot of
$\tau$ as a function of $\Delta_0$ with $\lambda= \pi$ is shown in
Fig.\ \ref{fig5}(d). We find that $\tau$ shows a steep rise around
$\Delta_0 \simeq \Omega,\delta$. This allows us to conclude that the
Floquet phases are stable for a very long timescale as long as we
are in the regime $\Delta_0 \gg \Omega,\delta$.

\section{Discussion}
\label{diss}

In this work, we have identified the Floquet phases of a
periodically driven Rydberg atom arrays. Such phases can be tuned as
a function of the drive frequency; our analysis identifies special
drive frequencies which satisfies $\Delta_0 T/\hbar = 4 m \pi$,
where $m$ is a positive integer, for a square pulse protocol and
$\Delta_0 T/\hbar= 2 \pi \eta_m$ for a cosine protocol. At these
drive frequencies, one finds density-wave ordered Floquet phases. In
the high drive amplitude regime, we find a large prethermal
timescale for which these Floquet phases are stable and are
accurately derived by the analytical first-order Floquet Hamiltonian
$H_F^{(1)}$ derived in Sec.\ \ref{analytic}. We note here that
although we have carried all the numerics using the square-pulse
protocol, the results of Sec.\ \ref{analytic} strongly suggest that
analogous phenomenon exists for continuous drive protocols; the
expression for the special frequencies for the cosine protocol is
given by Eq.\ \ref{magf}. We have also presented a method to detect
these Floquet phases and the transitions between them via
measurement of the equal-time density-density correlation function
$C_3(nT)$. We note that the Floquet phases, unlike their
thermodynamic counterparts, are not readily accessible in
experiments; our results therefore provide a useful experimental
tool for detection of these phases.

\begin{figure}
\rotatebox{0}{\includegraphics*[width= \linewidth]{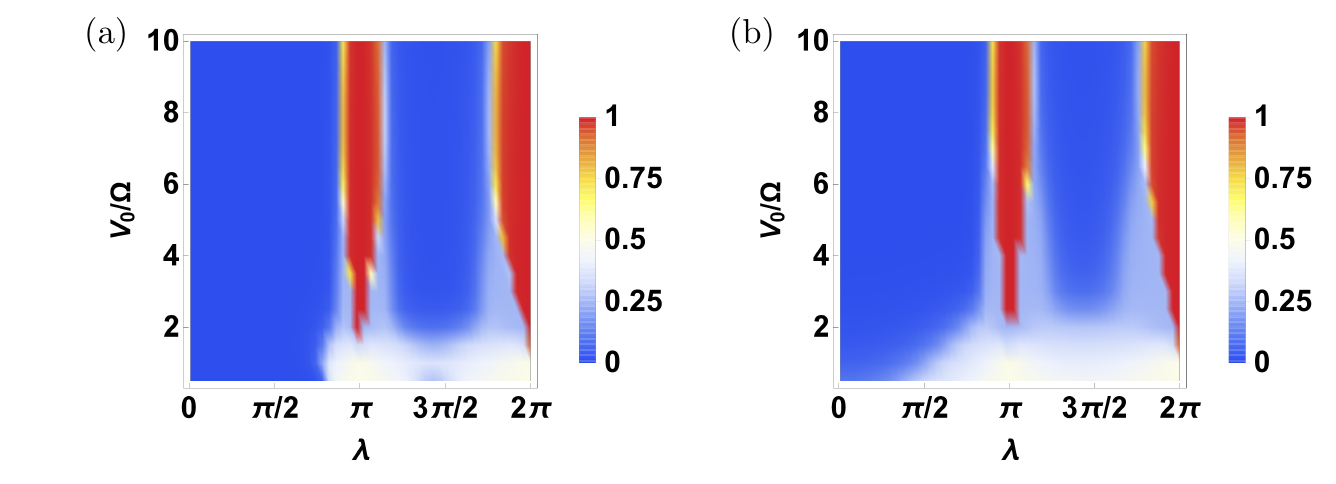}}
\caption{Plot of $O_0$ as a function of $V_0$ and $\lambda$ showing
the Floquet phases obtained using ED starting from (a)
$H_{Fs}^{(1)}$ (Eq.\ \ref{hf1}) and keeping the full Hilbert space
where neighboring Rydberg excitations are allowed and (b) $H_p(t)$
(Eq.\ \ref{ham2}) with the same square pulse protocol but working
with the projected Hilbert space with no nearest-neighbor Rydberg
excitations. For both plots $\delta=0.75 \Omega$, $\Delta_0=100
\Omega$, and $L_x=L_y=4$.} \label{fig6}
\end{figure}

In the previous sections, we have used the approximation of large
$V_0$ for obtaining these phases. This is not an essential feature
of our analysis as can be seen by comparing Eqs.\ \ref{hf1} and
\ref{projfl}. The first of these (Eq.\ \ref{hf1}) obtains the
Floquet Hamiltonian without any additional approximation for $V_0$
while the second is derived in the large $V_0$ regime. Both these
Floquet Hamiltonians provided identical expressions for the special
frequencies $\omega_m^{\ast}$ (Eq.\ \ref{magf}). The reason for
choosing the latter when it comes to exact numerics is that it has a
smaller Hilbert space which allows access to larger system sizes for
carrying out ED. To ascertain this fact, we show a comparison in
Fig.\ \ref{fig6} between the Floquet phases obtained by applying ED
on $H_{Fs}^{(1)}$ keeping the full Hilbert space and that obtained
by diagonalizing $U$ within the constrained Hilbert space. The
result of the phase diagram obtained from $H_{FS}^{(1)}$ is given in
Fig.\ \ref{fig6}(a). Fig. \ref{fig6}(b) shows the Floquet phases
obtained using the projected subspace for $N=16$ sites; the phase
diagram is similar to the one obtained for $N=24$ sites (Fig.\
\ref{fig2}(b)). A comparison of this phase diagram with the one
shown in Fig.\ \ref{fig6}(a) shows that they differ qualitatively
only for $V_0 \le \delta, \Omega$ and $\Delta_0 T/\hbar \ll \pi$;
the spurious interpolating behavior obtained using the projected
Hilbert space does not appear in this regime and is replaced by the
disordered phase in the exact phase diagram. However, in other
regimes, there is excellent agreement between the two phase diagrams
including around $V_0 \simeq \delta$ when $\lambda \ge \pi$. The
last feature owes its existence to the reduction of $\Omega_{\rm
eff} \sim \sin \lambda/\lambda$ in this regime which is equivalent
to an effective increase in $V_0$.

Finally, we discuss experiments which can test our theory. We
propose a standard experimental setup involving Rydberg atoms in a
rectangular array where the detuning of these atoms are changed
periodically with time according to either a square pulse or a
cosine protocol. We predict the existence of special frequencies
where the system should exhibit a star ordered Floquet phase at large
$V_0$. Such a phase would leave its imprint on the time evolution of
the correlation function $C_3$ starting from an initial Fock state
with star order; in the ordered phase $C_3$ will be very nearly time
independent. The transition between the star and the disordered
phase can be achieved by tuning the drive frequency; such a
transition will be reflected in the behavior of $C_3$ as discussed
in Sec.\ \ref{corr}.

In conclusion, we have discussed the Floquet phases of Rydberg atoms
arranged in a rectangular array. We have provided a way of
experimentally detecting of these phases and the transitions between
them via measurements of equal-time correlation functions; moreover, we
have identified the high drive amplitude regime where such phases
are stable over a long prethermal time scale. Within this time
scale, their properties can be described by the first-order Floquet
Hamiltonian obtained using FPT.

\section{Acknowledgements}

S.G. acknowledges CSIR, India for support through Project No.
09/080(1133)/2019-EMR-I. D.S. thanks SERB, India for support through
project JBR/2020/000043. K.S. thanks SERB, India for support through
project JCB/2021/000030.

\end{document}